\documentclass{emulateapj}
\lefthead{GAUDI ET AL.} \righthead{ROTATION PERIOD OF SEDNA}

\def\eq#1{equation (\ref{#1})}

\def\au{\rm AU}
\def\hours{{\rm hours}}
\def\days{{\rm days}}

\begin{document}

\title{On the Rotation Period of (90377) Sedna} 
\author{B.\ Scott Gaudi, Krzysztof Z.\ Stanek,  Joel D.\ Hartman, Matthew J.\ Holman, and Brian A.\ McLeod}

\affil{Harvard-Smithsonian Center for Astrophysics, 60 Garden St., Cambridge, MA 02138}
\email{sgaudi@cfa.harvard.edu}

\begin{abstract}
We present precise, $\sim 1\%$, $r$-band relative photometry of the
unusual solar system object (90377) Sedna.  Our data consist of 143
data points taken over eight nights in October 2004 and
January 2005.  The RMS variability over the longest contiguous stretch
of five nights of data spanning nine days is only $\sim 1.3\%$.  This
subset of data alone constrain the amplitude of any long-period
variations with period $P$ to be $A\la 1\%(P/20~\days)^2$.  Over the
course of any given $\sim 5$-hour segment, the data exhibits
significant linear trends not seen in a comparison star of similar
magnitude, and in a few cases these segments show clear evidence for
curvature at the level of a few millimagnitudes per hour$^2$.  These
properties imply that the rotation period of Sedna is $O(10~\hours)$,
cannot be $\la 5~\hours$, and cannot be $\ga 10~\days$, unless the
intrinsic light curve has significant and comparable power on multiple
timescales, which is unlikely.  A sinusoidal fit
yields a period of
$P=(10.273\pm0.002)\hours$ and semi-amplitude of $A=(1.1 \pm 0.1)\%$.
There are additional acceptable fits with flanking periods separated by
$\sim 3~{\rm minutes}$, as well as another class of fits with $P\sim
18~\hours$, although these later fits appear less viable based on
visual inspection.
Our results indicate that the period of Sedna is likely consistent
with typical rotation periods of solar system objects, thus obviating
the need for a massive companion to slow its rotation.
\end{abstract}

\section{Introduction}\label{sec:intro}

There is increasing evidence for the existence of an extended
scattered disk; a massive population of objects orbiting beyond the
Kuiper belt \citep{gladman02}.  These objects have orbits with
substantial eccentricities and inclinations and are distinct from
Kuiper Belt Objects (KBOs) in that their perihelia are little affected
by gravitational perturbations from Neptune.  Thus it appears that 
Neptune cannot be responsible for
their unusual orbits, and several novel mechanisms to explain the
origin of these object have been proposed \citep{ml04,kb04,stern05}.
The total mass in these objects is poorly known because only a handful
of members have been discovered.  These include the recently detected
object (90377) Sedna ($=2003~{\rm VB}_{12}$), whose orbit has a
semimajor axis of $a \simeq 500~\au$ and a perihelion of $q \simeq
80~\au$ \citep{sedna}.

Sedna appears to be extreme in several ways in addition to its unusual
orbit.  It is intrinsically bright, with an absolute magnitude of
$H=1.6$, implying that it is one of the largest known minor planets.  
Unpublished reports also indicate
that it is quite red, has a relatively high albedo, a weak opposition
surge, and has a very long rotation period, with $P\sim
20-40~\days$ \citep{brown04b}.  The latter claim is especially interesting in light of
the fact that a {\it Hubble Space Telescope} snapshot of Sedna
revealed no evidence for a large companion that could have tidally decreased
Sedna's rotation period from typical solar system rotation
periods of $O(10~\hours)$ to a longer period of $\sim 20~\days$.

Here we present precise relative photometry of Sedna that
indicates a rotation period of $O(10~\hours)$, and rules out
rotation periods longer than $\sim 10~\days$, under reasonable assumptions.
The rotation period of Sedna is likely within the range
of typical solar system objects, obviating the need for a massive companion.

\begin{deluxetable*}{ccccc}
\tablecaption{\sc Sedna Relative Photometry and Phase}
\tablewidth{0pt}
\tabletypesize{\scriptsize}
\tablehead{
  \colhead{Date} &
  \colhead{HJD-2450000.} &
  \colhead{$\Delta r$\tablenotemark{a}} &
  \colhead{$\sigma_{\Delta r}$} &
  \colhead{Phase Angle ($^\circ$)} 
}
\startdata
UT 2004 Oct 8 
&3286.83028 & -0.001 & 0.005 & 0.3759\\
&3286.84411 & -0.004 & 0.006 & 0.3758\\
&3286.84823 & -0.013 & 0.006 & 0.3757\\
&3286.85249 & -0.001 & 0.007 & 0.3757\\
&3286.85723 & -0.001 & 0.007 & 0.3757\\
&3286.86124 &  0.005 & 0.009 & 0.3757\\
&3286.86525 & -0.010 & 0.007 & 0.3756\\

\enddata
\tablecomments{[The complete version of this table is in the
electronic edition of the Journal.  The printed edition contains only
a sample.]}
\tablenotetext{a}{Note that the photometry has
an arbitrary zero point which differs for the data
taken during UT 2004 Oct 8-9, UT 2004 Oct 16, and UT 2005 Jan 7-15.}
\label{tab:data}
\end{deluxetable*}

\begin{figure}
\epsscale{1.8} 
\plotone{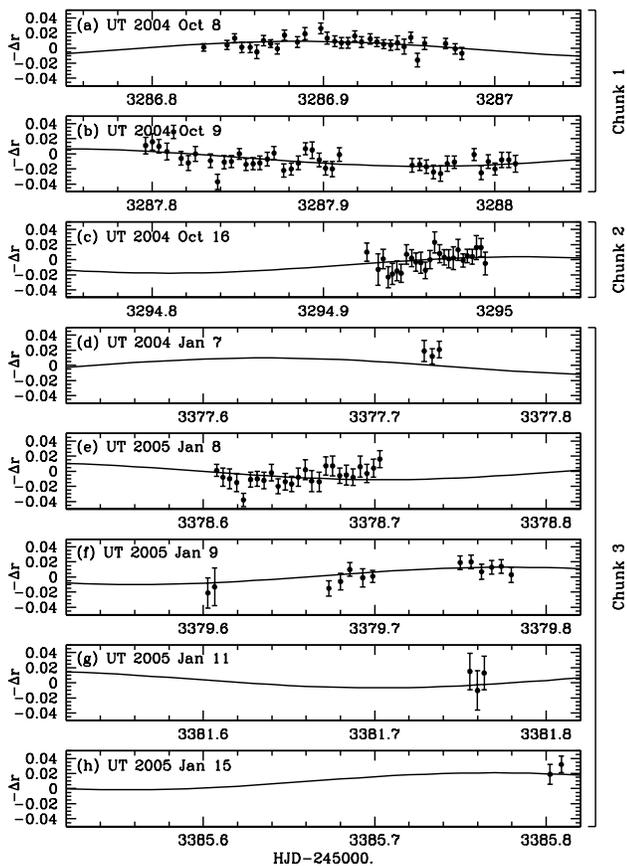}
\caption{\label{fig:one} 
Relative photometry of Sedna.  The data points show the $r$-magnitude
of Sedna versus HJD-2450000., relative to an arbitrary offset that is
independent for each chunk.  The solid line shows the best-fit
sinusoidal model.  Individual panels show data from
the nights of (a) UT 2004 Oct 8, (b) UT 2004 Oct 9, (c) UT
2005 Oct 16,  (d) UT 2004 Jan 7, (e) UT 2004 Jan 8, (f) UT 2004 Jan 9, (g) UT 2004 Jan 11, (h) UT 2004 Jan 15.
}
\end{figure}

\section{Observations and Data Reduction}\label{sec:data}

We observed Sedna over eight nights in October 2004
(UT 2004 Oct 8,9,16) and January 2005 (UT 2005 Jan 7-9,11,15).  Photometric
data were obtained with the MegaCam CCD camera \citep{mcleod00}
on the MMT 6.5m telescope.  The MegaCam instrument uses 36
2048x4608 CCDs to cover a 24'x24' field-of-view with a pixel scale of
0.08''.  Our primary science goal was to search for small KBOs, 
but we chose to target the field of Sedna to simultaneously
acquire a precise light curve for this unusual object.  The results of
the KBO search will be presented elsewhere.  Conditions during the
observations ranged from good to poor, with image FWHMs in the range
0.7-1.9''.  All data were taken with a Sloan $r$-band filter with 2x2 image binning.
Exposure times were 300-450 seconds.  The apparent motion of Sedna during 
our observations was $\sim 1''/{\rm hr}$, so trailing losses are negligible.

The images were further binned and then reduced in the usual manner.
Photometry was performed in two ways: using PSF-fitting
photometry with the DAOPHOT II package \citep{stetson87,stetson92},
and using image-subtraction photometry with the ISIS 2.1 package
\citep{al98,alard00}.  For the DAOPHOT reductions, relative photometry
of Sedna was derived using 10-50 reference stars.  

For moving objects, one must take care to consider background stars or
galaxies that may be blended with the target in only a subset of
exposures, potentially leading to artificial variability when using
PSF-fitting photometry.  In fact, during the night of UT 2004 Jan 8,
Sedna was blended with a background object that was $\sim 3.5$
magnitudes fainter.  Image subtraction photometry
eliminates any constant, stationary objects, and so removes such
contamination.  On the other hand, the quality of PSF-fitting
photometry can be comparable to image-subtraction photometry for
uncontaminated objects in relatively sparse fields.  Furthermore, we
have found that DAOPHOT can extract reliable measurements from very
poor-quality frames, where ISIS fails.  Therefore, in order to provide the best
possible photometry, we adopted a hybrid approach, combining
PSF-fitting photometry for the nights which showed no evidence for
contaminating
background objects (UT 2004 Oct 8-9,16 and UT 2005 Jan 11), and
image-subtraction photometry for the remainder of the nights (UT 2005
Jan 7-9,15).  We stress that, for nights with no contamination, the
light curves produced by the two methods are completely consistent.
We used the DAOPHOT-reported errors for all data, as we judged these
to be more reliable than ISIS-reported errors.

Due to Sedna's proper motion, it was not possible to use the same
reference stars or images and thus tie the photometry to the same zero
point for the entire dataset.  Therefore the data consist of three
`chunks', corresponding to data taken on UT 2004 Oct 8-9, UT 2004 Oct
16, and UT 2005 Jan 7-15.  Each of these chunks have an independent
zero point.  Although the relative offset and absolute photometric
calibration of these chunks could be
determined by various methods, these methods all require additional data.
These data are not currently available.   We therefore chose to
present only relative photometry.  This final photometry, consisting
of 143 data points, is listed in Table 1, where we have subtracted the
error-weighted mean instrumental magnitude from each chunk.  We note
that the apparent magnitude of Sedna during our observations was
approximately $R\sim 21$.

\section{Analysis}\label{sec:ana}

Figure \ref{fig:one} shows the light curve for Sedna, where each panel
corresponds to a different night.  The nights belonging to the three
separate chunks are indicated; each chunk has an independent zero
point.  The solid curve is a sinusoidal model, which is
described below.

Several relatively model-independent conclusions can be drawn from the
properties of the light curve.  First, the RMS deviation during the largest chunk spanning
nine nights during UT 2005 Jan 7-15 is only $\sim 1.3\%$.  In addition,
these data show no evidence for significant curvature;
a simple second-order polynomial fit to the January data yields an upper limit
to the coefficient of the quadratic term of $c_2\le 440~\mu{\rm
mag/day^2}$.  This implies that if the light
curve amplitude is large, the rotation period must be long. For example,
for a sinusoidal light curve, this corresponds a
limit on the semi-amplitude of $A \la c_2P^2/2\pi^2 \sim 1\% (P/20~\days)^2$ for
large $P$.  Second, the
data during any given individual night spanning $\la 5~\hours$ generally have very small RMS
deviations.  For example, the RMS for the night of UT 2004 Oct 8 is
only 0.7\%.  Nevertheless, several nights show evidence for
significant variability that is not seen in a comparison star of
similar magnitude.  In many cases, this variability is
consistent with a simple linear trend, which argues that the period
cannot be $\la 5~\hours$.
However, for a few nights, curvature is evident.  For example, a second-order polynomial fit to
the UT 2004 Oct 8 data yields a $\sim 3\sigma$ detection of curvature
with $c_2=(-3.8\pm 1.2){\rm millimag/hr^2}$.  Similarly, a fit to the UT
2004 Oct 9 data yields $c_2=(2.0\pm 0.9){\rm millimag/hr^2}$.  
The detection
of significant curvature, the fact that the curvature
on adjacent nights has opposite sign, and the fact that the difference in
mean magnitudes between adjacent nights is $\sim 1\%$, argue 
that the period must be $O(10~\hours)$.  This assumes that the primary power in the intrinsic 
light curve occurs at only one period.  We believe this is a
reasonable assumption.
 
\begin{deluxetable*}{cccccccc}
\tablecaption{\sc  Fit Parameters}
%\tablewidth{0pt}
\tabletypesize{\scriptsize}
\tablehead{
  \colhead{$P$} &
  \colhead{$A$} &
  \colhead{$\phi_0$} &
  \colhead{$k$} &
  \colhead{$F_{0,1}$} &
  \colhead{$F_{0,2}$} &
  \colhead{$F_{0,3}$} &
  \colhead{$\chi^2$} \\
 \colhead{(hours)} &
  \colhead{} &
  \colhead{} &
  \colhead{(${\rm deg}^{-1}$)} &
  \colhead{} &
  \colhead{} &
  \colhead{} &
  \colhead{(136 dof) } 
}
\startdata
    10.273$\pm$     0.002  &  0.011$\pm$0.001  &   0.73$\pm$0.12  &   0.2$\pm$0.2  & 1.004$\pm$0.001  & 1.015$\pm$0.002  & 0.970$\pm$0.002  &  150.0\\
    10.321$\pm$     0.002  &  0.010$\pm$0.001  &   5.60$\pm$0.12  &   0.2$\pm$0.2  & 1.004$\pm$0.001  & 1.019$\pm$0.003  & 0.969$\pm$0.002  &  150.0\\
    17.991$\pm$     0.006  &  0.011$\pm$0.002  &   4.43$\pm$0.17  &   0.2$\pm$0.2  & 1.001$\pm$0.008  & 1.016$\pm$0.025  & 0.978$\pm$0.036  &  150.0\\
    17.845$\pm$     0.006  &  0.011$\pm$0.002  &   5.84$\pm$0.16  &   0.2$\pm$0.2  & 1.002$\pm$0.008  & 1.011$\pm$0.025  & 0.977$\pm$0.036  &  150.2\\
    18.139$\pm$     0.006  &  0.010$\pm$0.002  &   3.01$\pm$0.18  &   0.2$\pm$0.2  & 1.003$\pm$0.008  & 1.026$\pm$0.025  & 0.971$\pm$0.036  &  150.9\\
\enddata
\label{tab:fits}
\end{deluxetable*}
\bigskip

\begin{figure}
\epsscale{1.0} 
\plotone{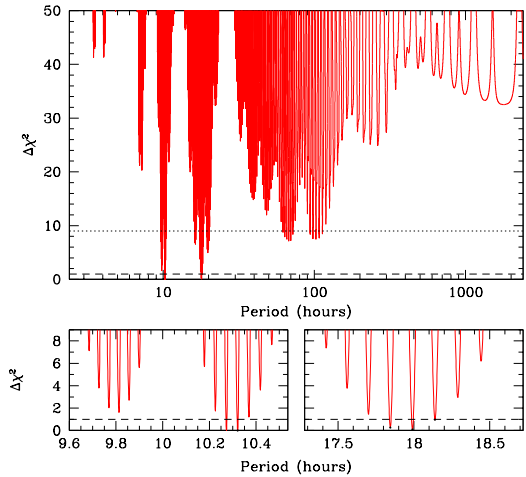}
\caption{\label{fig:two} 
The difference in $\chi^2$ of a sinusoidal model fit
to the Sedna light curve from the minimum 
$\chi^2$ of the best-fit model with $P=(10.273\pm0.002)\hours$, 
as a function of the period of the model. The top
panel shows the full range of periods searched, whereas the bottom
panels show close-ups of the two most significant classes of fits.
The horizontal lines show $\Delta\chi^2=1$ (dashed) and $9$ (dotted).
}
\end{figure}

\begin{figure}
\epsscale{1.0} 
\plotone{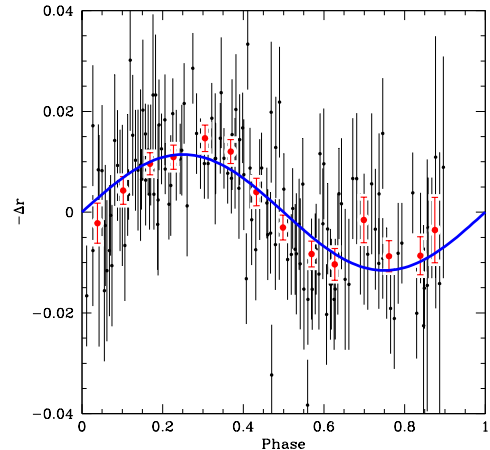}
\caption{\label{fig:three} 
The black points show the relative photometry of Sedna,
phased to the best-fit period of $P=(10.273\pm 0.002)~\hours$, with the fitted zero point
and phase variations subtracted.  The red points show the data binned
into intervals of $0.067$ in phase.  The blue curve shows the best-fit 
sinusoidal model.
}
\end{figure}

\begin{figure}
\epsscale{1.0} 
\plotone{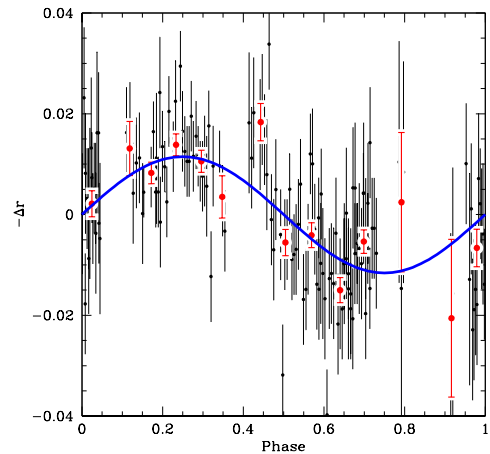}
\caption{\label{fig:four} 
Same as Figure \ref{fig:three}, except for the best model
of the second class of acceptable fits, with
a period of $P=(17.991 \pm0.006)~\hours$. Although the 
$\chi^2$ of this model for the unbinned data is 
nearly indentical to that of the model with
$P=(10.273\pm 0.002)~\hours$ shown in Figure \ref{fig:three},
the $\chi^2$ of the binned data
is considerably worse. Thus the model with 
$P\sim 10~\hours$ is favored.
}
\end{figure}

We fit the light curve to the seven-parameter model,
\begin{equation}
F(t_i)=A\sin\left[ \frac{2\pi}{P}(t_i-t_0)+\phi_0\right]-k[\alpha(t_i)-\alpha_0]+F_{0,j},
\label{eqn:model}
\end{equation}
where $F(t_i)$ is the flux at the time $t_i$ of observation $i$, $\alpha$ is the phase
angle of Sedna at this time, $k$ is the coefficient of the phase function\footnote{We have
assumed a linear phase function. This is appropriate given the relatively small range of 
phase angles spanned by our dataset \citep{bowell89}. See Table \ref{tab:data}.},
$F_{0,j}$ is the flux zero point for chunk $j$, and  
$t_0-2450000.=3308.23289$ and $\alpha_0=0.4039$ are the error-weighted mean
observation times and phase angles, respectively.  Note that
we are fitting relative photometry, and thus $A$, $k$, and $F_{0,j}$ 
are dimensionless.  In practice, we
expand the sinusoidal term in \eq{eqn:model} into separate sine and
cosine terms, and then perform a linear fit in flux to the
coefficients of these terms, the phase angle term, and the constant
terms.  We then reconstruct the more physical parameters $A$ and
$\phi_0$ from the coefficients of the sine and cosine terms.  This has
the advantages that the only non-linear variable that must be fitted
is $P$, and errors on the parameters $A,\phi_0,k$, and $F_{0,j}$ can
be determined analytically at fixed $P$.  We constrain $k$ to be
within $1\sigma$ of the range $0\le k \le 0.3~{{\rm deg}^{-1}}$, although the
exact form of the contraint has little effect on the results.   Note
that, aside from the phase angle term, \eq{eqn:model} is equivalent to
a Lomb-Scargle periodogram with a floating mean
\citep{lomb76,scargle82,cumming04}.

We search for fits in the range $-1\le \log(P/{\rm day})\le 3$, with steps of
$\delta P/P=4\times 10^{-6}$.  The resulting periodogram, here
displayed as $\Delta\chi^2\equiv \chi^2-\chi^2_{\rm min}$ versus $P$, is
shown in Figure \ref{fig:two}. The best-fit has $\chi^2_{\rm
min}=150.0$ for $143-7=136~{\rm dof}$, indicating a good fit.  For reference, a
constant flux fit to the data yields $\chi^2=272.4$ for $140~{\rm dof}$.  Thus the
detection of variability, as judged by the improvement in $\chi^2$, is extremely
significant. 
The parameters for the fit are $P=(10.273\pm0.002)\hours$ and $A=(1.1\pm0.1)\%$.
The phase angle coefficient $k$ is poorly constrained, due to the fact
that the separate chunks are not tied together, and thus the time
baseline for determining $k$ is limited to the $\sim 9$ day span of our January data.
Figure \ref{fig:three} shows the light curve phased to the 
best-fit period, with the constant flux and phase angle terms subtracted, along
with the model fit.  
The model appears to describe the data 
reasonably well. 

Flanking the best-fit period are additional fits separated by $\sim
2.82~{\rm minutes}$ (see Fig.\ \ref{fig:two}); these correspond to fits in which there are one or more
additional cycles between the October and January datasets, i.e.\
where $P_1^{-1}-P_2^{-1}\simeq \pm n (90~\days)^{-1}$ for integer $n$.
In addition, there is a cluster of fits that is separated by 
$\sim 27.6~{\rm minutes}$ from the best-fit period. These corresponds
to fits in which there is one additional cycle between UT Oct 9 and 
UT Oct 16.  Finally, there are also diurnal aliases near $P\simeq
18~\hours$ and $3~\days$ (and the associated aliases of these
aliases).  Fits near the latter period are allowed at the $\sim 3\sigma$ level.

We find a total of five fits that are statistically indistinguishable 
($\Delta\chi^2 \la 1$) from the best fit.  The parameters of these
fits are given in Table 2.  Two of these fits have $P\sim 10~\hours$,
and appear equally good by eye.  The other three fits have
$P\simeq 18~\hours$.  Although these fits are statistically acceptable, they appear
much less convincing upon inspection of the phased light curves, one
example of which is shown in Figure \ref{fig:four}.
The amplitude is relatively constant for all the acceptable fits, with $A\simeq 1\%$.
Models with $P\ge 10~\days$ are ruled out at the $> 4\sigma$ level. 
Refitting the data after subtracting the flux predicted
by the best-fit model reveals no  significant additional
periodicities.  

As a sanity check, we repeated the analysis described above on a light
curve constructed from comparison stars of similar magnitude as Sedna.
We find no evidence for variability at the level exhibited by Sedna.  The
best fit has an improvement in $\chi^2$ over a constant flux
model of $\sim 38$ for 4 additional degrees of freedom, with an amplitude
of only  $A=(0.38\pm 0.06)\%$.

\section{Summary and Discussion}\label{sec:disc}

We have presented relative photometry of the unusual solar system object
Sedna, obtained with the MMT 6.5m telescope over eight nights in two campaigns 
in October 2004 and January 2005.
The light curve during the longest contiguous stretch of nine days has a remarkably
small RMS of $\sim 1.3\%$, and exhibits no significant curvature, which
severely constrains the amplitude of any long-term variability to
$A\le 1\% (P/20~\days)^2$.
The light curve during any individual night exhibits significant variability
that is not seen in a  comparison star of similar brightness.  The photometry 
from several individual nights shows significant curvature over the span of $\sim 5$ hours.

These properties indicate that the period of Sedna is $O(10~\hours)$,
and cannot be larger than $\sim 10~{\rm days}$.  
A sinusoidal model fit to Sedna yields a best-fit period of
$P=(10.273\pm0.003)\hours$ and semi-amplitude $A=(1.0 \pm 0.1)\%$,
with additional acceptable fits with flanking periods separated by
$\sim 3~{\rm minutes}$, as well as another class of fits with $P\sim
18~\hours$, although these later fits appear less viable based on
visual inspection.  We note that, if the variability is due to an
aspherical shape such as a triaxial ellipsoid, the true rotation
period is twice the fitted period.  There also exist fits at the
diurnal aliases of the primary period with $P\sim 3~\days$ that are
marginally acceptable at the $3\sigma$ level. Fits with $P\le
10~\hours$ or $P\ge 10~\days$ are ruled out at the $\ge 3\sigma$
level.  Thus we conclude that the rotation period of Sedna is most
likely $P\sim 10~\hours$, although other periods cannot be completely
excluded.  Additional observations
should be pursued to distinguish between
the various viable fits found here, and so firmly identify the true
rotation period of Sedna.
The best-fit rotation period of $\sim 10~\hours$ makes
Sedna entirely typical of the bulk of solar system objects, including
main-belt asteroids \citep{ph00,harris02}, as well as the $\sim$dozen
KBOs with measured rotation rates \citep{sheppard02}.

We conclude that there is no real evidence that the period of Sedna is
extraordinarily long ($P\ge 10~\days$) or even unusual.  Therefore there is
no compelling reason to invoke a massive companion to spin down
Sedna's rotation period.

\acknowledgments 
BSG was supported by a Menzel Fellowship from the Harvard College
Observatory. KZS acknowledges support from the William F.\ Milton
Fund. We would like to thank Roman Rafikov for helpful discussions,
Scott Kenyon for reading the manuscript, and Perry Berlind, Emeric Le
Floc'h, Casey Papovich, Jane Rigby, and Kurtis Williams for assistance
in acquiring additional data.

\clearpage

\LongTables

\begin{deluxetable}{ccccc}
\tablecaption{\sc Sedna Relative Photometry and Phase (Full Table)}
\tablewidth{0pt}
\tabletypesize{\scriptsize}
\tablehead{
  \colhead{Date} &
  \colhead{HJD-2450000.} &
  \colhead{$\Delta r$\tablenotemark{a}} &
  \colhead{$\sigma_{\Delta r}$} &
  \colhead{Phase Angle ($^\circ$)} 
}
\startdata
UT 2004 Oct 8 
& 3286.83028 & -0.001 & 0.005 & 0.3759 \\
& 3286.84411 & -0.004 & 0.006 & 0.3758 \\
& 3286.84823 & -0.013 & 0.006 & 0.3757 \\
& 3286.85249 & -0.001 & 0.007 & 0.3757 \\
& 3286.85723 & -0.001 & 0.007 & 0.3757 \\
& 3286.86124 & 0.005 & 0.009 & 0.3757 \\
& 3286.86525 & -0.010 & 0.007 & 0.3756 \\
& 3286.86923 & -0.006 & 0.005 & 0.3756 \\
& 3286.87324 & 0.001 & 0.007 & 0.3756 \\
& 3286.87721 & -0.017 & 0.007 & 0.3755 \\
& 3286.88508 & -0.008 & 0.007 & 0.3755 \\
& 3286.88937 & -0.019 & 0.008 & 0.3754 \\
& 3286.89847 & -0.026 & 0.007 & 0.3754 \\
& 3286.90247 & -0.013 & 0.008 & 0.3753 \\
& 3286.90671 & -0.009 & 0.007 & 0.3753 \\
& 3286.91076 & -0.007 & 0.007 & 0.3752 \\
& 3286.91472 & -0.007 & 0.007 & 0.3752 \\
& 3286.91867 & -0.016 & 0.007 & 0.3752 \\
& 3286.92265 & -0.008 & 0.007 & 0.3751 \\
& 3286.92768 & -0.012 & 0.007 & 0.3751 \\
& 3286.93158 & -0.008 & 0.006 & 0.3750 \\
& 3286.93554 & -0.005 & 0.006 & 0.3750 \\
& 3286.93950 & -0.004 & 0.007 & 0.3749 \\
& 3286.94340 & -0.007 & 0.009 & 0.3749 \\
& 3286.94738 & -0.002 & 0.010 & 0.3749 \\
& 3286.95137 & -0.014 & 0.007 & 0.3748 \\
& 3286.95530 & 0.016 & 0.009 & 0.3748 \\
& 3286.95929 & -0.006 & 0.008 & 0.3748 \\
& 3286.97152 & -0.006 & 0.007 & 0.3747 \\
& 3286.97705 & 0.001 & 0.007 & 0.3746 \\
& 3286.98105 & 0.007 & 0.008 & 0.3746 \\
UT 2004 Oct 9
& 3287.79633 & -0.011 & 0.011 & 0.3676 \\
& 3287.80032 & -0.016 & 0.011 & 0.3676 \\
& 3287.80436 & -0.010 & 0.009 & 0.3676 \\
& 3287.80875 & -0.003 & 0.011 & 0.3675 \\
& 3287.81288 & -0.029 & 0.009 & 0.3675 \\
& 3287.81709 & 0.006 & 0.009 & 0.3674 \\
& 3287.82135 & 0.012 & 0.010 & 0.3674 \\
& 3287.82553 & 0.000 & 0.010 & 0.3674 \\
& 3287.83422 & 0.009 & 0.009 & 0.3673 \\
& 3287.83827 & 0.037 & 0.010 & 0.3672 \\
& 3287.84231 & 0.011 & 0.009 & 0.3672 \\
& 3287.84635 & 0.010 & 0.008 & 0.3672 \\
& 3287.85118 & 0.000 & 0.007 & 0.3671 \\
& 3287.85518 & 0.014 & 0.008 & 0.3671 \\
& 3287.85914 & 0.013 & 0.008 & 0.3671 \\
& 3287.86317 & 0.012 & 0.009 & 0.3670 \\
& 3287.86728 & 0.007 & 0.009 & 0.3670 \\
& 3287.87132 & -0.001 & 0.009 & 0.3670 \\
& 3287.87678 & 0.022 & 0.009 & 0.3669 \\
& 3287.88130 & 0.020 & 0.008 & 0.3669 \\
& 3287.88542 & 0.012 & 0.009 & 0.3669 \\
& 3287.88954 & -0.007 & 0.008 & 0.3668 \\
& 3287.89360 & -0.005 & 0.011 & 0.3668 \\
& 3287.89764 & 0.008 & 0.009 & 0.3668 \\
& 3287.90167 & 0.019 & 0.009 & 0.3667 \\
& 3287.90569 & 0.020 & 0.010 & 0.3667 \\
& 3287.90969 & 0.001 & 0.009 & 0.3667 \\
& 3287.95197 & 0.015 & 0.009 & 0.3662 \\
& 3287.95628 & 0.014 & 0.008 & 0.3662 \\
& 3287.96039 & 0.017 & 0.009 & 0.3662 \\
& 3287.96446 & 0.024 & 0.009 & 0.3661 \\
& 3287.96847 & 0.026 & 0.010 & 0.3661 \\
& 3287.97248 & 0.013 & 0.010 & 0.3661 \\
& 3287.97643 & 0.011 & 0.008 & 0.3660 \\
& 3287.98805 & 0.001 & 0.008 & 0.3659 \\
& 3287.99222 & 0.025 & 0.009 & 0.3659 \\
& 3287.99626 & 0.010 & 0.009 & 0.3659 \\
& 3288.00030 & 0.020 & 0.008 & 0.3659 \\
& 3288.00428 & 0.008 & 0.010 & 0.3658 \\
& 3288.00826 & 0.008 & 0.010 & 0.3658 \\
& 3288.01231 & 0.013 & 0.011 & 0.3658 \\
UT 2004 Oct 16
& 3294.92551 & -0.010 & 0.012 & 0.3049 \\
& 3294.93194 & 0.013 & 0.021 & 0.3048 \\
& 3294.93511 & -0.001 & 0.013 & 0.3048 \\
& 3294.93771 & 0.023 & 0.014 & 0.3048 \\
& 3294.94036 & 0.019 & 0.014 & 0.3048 \\
& 3294.94297 & 0.015 & 0.013 & 0.3047 \\
& 3294.94560 & 0.018 & 0.012 & 0.3047 \\
& 3294.94859 & -0.007 & 0.013 & 0.3047 \\
& 3294.95159 & -0.002 & 0.011 & 0.3047 \\
& 3294.95417 & 0.003 & 0.012 & 0.3046 \\
& 3294.95682 & 0.004 & 0.014 & 0.3046 \\
& 3294.95942 & 0.014 & 0.011 & 0.3046 \\
& 3294.96230 & 0.000 & 0.012 & 0.3045 \\
& 3294.96511 & -0.023 & 0.014 & 0.3045 \\
& 3294.96771 & -0.008 & 0.012 & 0.3045 \\
& 3294.97066 & -0.003 & 0.010 & 0.3045 \\
& 3294.97333 & -0.001 & 0.012 & 0.3044 \\
& 3294.97592 & -0.002 & 0.015 & 0.3044 \\
& 3294.97858 & -0.013 & 0.014 & 0.3044 \\
& 3294.98151 & 0.001 & 0.009 & 0.3044 \\
& 3294.98411 & -0.005 & 0.009 & 0.3043 \\
& 3294.98672 & -0.004 & 0.010 & 0.3043 \\
& 3294.98936 & -0.016 & 0.016 & 0.3043 \\
& 3294.99201 & -0.016 & 0.012 & 0.3043 \\
& 3294.99467 &  0.005 & 0.015 & 0.3042 \\
UT 2005 Jan 7
& 3377.72877 & -0.019 & 0.014 & 0.5415 \\
& 3377.73359 & -0.012 & 0.010 & 0.5415 \\
& 3377.73765 & -0.021 & 0.011 & 0.5415 \\
UT 2005 Jan 8
& 3378.60779 & -0.001 & 0.008 & 0.5463 \\
& 3378.61168 &  0.008 & 0.012 & 0.5463 \\
& 3378.61564 &  0.010 & 0.013 & 0.5463 \\
& 3378.61955 &  0.015 & 0.012 & 0.5464 \\
& 3378.62348 &  0.038 & 0.009 & 0.5464 \\
& 3378.62764 &  0.011 & 0.010 & 0.5464 \\
& 3378.63155 &  0.010 & 0.010 & 0.5464 \\
& 3378.63556 &  0.012 & 0.011 & 0.5464 \\
& 3378.63979 &  0.002 & 0.011 & 0.5465 \\
& 3378.64371 &  0.020 & 0.010 & 0.5465 \\
& 3378.64770 &  0.014 & 0.011 & 0.5465 \\
& 3378.65161 &  0.017 & 0.010 & 0.5465 \\
& 3378.65551 &  0.008 & 0.012 & 0.5466 \\
& 3378.65941 & -0.002 & 0.013 & 0.5466 \\
& 3378.66333 &  0.013 & 0.014 & 0.5466 \\
& 3378.66742 &  0.014 & 0.013 & 0.5466 \\
& 3378.67147 & -0.007 & 0.012 & 0.5466 \\
& 3378.67559 & -0.007 & 0.013 & 0.5467 \\
& 3378.67951 &  0.006 & 0.010 & 0.5467 \\
& 3378.68344 &  0.005 & 0.013 & 0.5467 \\
& 3378.68732 &  0.008 & 0.011 & 0.5467 \\
& 3378.69146 & -0.006 & 0.014 & 0.5468 \\
& 3378.69543 &  0.003 & 0.012 & 0.5468 \\
& 3378.69936 & -0.004 & 0.012 & 0.5468 \\
& 3378.70325 & -0.016 & 0.011 & 0.5468 \\
UT 2005 Jan 9
& 3379.60272 &  0.021 & 0.020 & 0.5517 \\
& 3379.60662 &  0.013 & 0.025 & 0.5517 \\
& 3379.67326 &  0.015 & 0.010 & 0.5520 \\
& 3379.67991 &  0.006 & 0.011 & 0.5521 \\
& 3379.68560 & -0.010 & 0.009 & 0.5521 \\
& 3379.69299 &  0.001 & 0.012 & 0.5522 \\
& 3379.69874 & -0.001 & 0.008 & 0.5522 \\
& 3379.75010 & -0.019 & 0.009 & 0.5525 \\
& 3379.75624 & -0.020 & 0.009 & 0.5525 \\
& 3379.76249 & -0.007 & 0.010 & 0.5526 \\
& 3379.76823 & -0.013 & 0.009 & 0.5526 \\
& 3379.77411 & -0.014 & 0.009 & 0.5526 \\
& 3379.77987 & -0.003 & 0.010 & 0.5527 \\
UT 2005 Jan 11
& 3381.75575 & -0.015 & 0.024 & 0.5629 \\
& 3381.75992 &  0.010 & 0.026 & 0.5629 \\
& 3381.76399 & -0.013 & 0.022 & 0.5629 \\
UT 2005 Jan 15
& 3385.80239 & -0.019 & 0.013 & 0.5817 \\
& 3385.80874 & -0.032 & 0.011 & 0.5817 \\
\enddata
\tablenotetext{a}{Note that the photometry has
an arbitrary zero point which differs for the data
taken during UT 2004 Oct 8-9, UT 2004 Oct 16, and UT 2005 Jan 7-15.}
\label{tab:datafull}
\end{deluxetable}

\end{document}